\documentclass[journal]{IEEEtran}
\usepackage{cite}

\ifCLASSINFOpdf
\else
\usepackage[dvips]{graphicx}
\fi
\usepackage[cmex10]{amsmath}
\begin{document}
%
\title{On Max-SINR Receiver for HMT System over Doubly Dispersive Channel}
%
%
%

\author{Kui~Xu,
        ~Youyun~Xu,~\IEEEmembership{Senior Member,~IEEE,}
        ~Dongmei~Zhang~
        and~Wenfeng~Ma
\thanks{Copyright (c) 2013 IEEE. Personal use of this material is permitted. However, permission to use this material for any other purposes must be obtained from the IEEE by sending a request to pubs-permissions@ieee.org.

Part of the material in this paper will presented at the 2012 IEEE
GLOBECOM Proceedings. This work was supported by the Jiangsu
Province Natural Science Foundation under Grant (BK2012055,
BK2011002), the Young Scientists Pre-research Fund of PLAUST under
Grant (No.KYTYZLXY1211).
K.Xu, Y. Xu, D. Zhang is with the Institute of Communications Engineering, PLA University of Science and Technology, Nanjing 210007, China. e-mail: (lgdxxukui@126.com)}}

%
%

\markboth{IEEE Transactions On Vehicular Technology, Accepted For Publication}%
{Shell \MakeLowercase{\textit{et al.}}: Bare Demo of IEEEtran.cls
for Journals}
%



\maketitle

\begin{abstract}
In this paper, a novel receiver for Hexagonal Multicarrier
Transmission (HMT) system based on the maximizing
Signal-to-Interference-plus-Noise Ratio (Max-SINR) criterion is
proposed. Theoretical analyses show that there is a timing offset
between the prototype pulses of the proposed Max-SINR receiver and
the traditional projection receiver. Meanwhile, the timing offset
should be matched to the channel scattering factor of the doubly
dispersive (DD) channel. The closed form timing offset expressions
of the prototype pulse for Max-SINR HMT receiver over DD channel
with different channel scattering functions are derived. Simulation
results show that the proposed Max-SINR receiver outperforms
traditional projection scheme and obtains an approximation to the
theoretical upper bound SINR performance. Consistent with the SINR
performance improvement, the bit error rate (BER) performance of HMT
system has also been further improved by using the proposed Max-SINR
receiver. Meanwhile, the SINR performance of the proposed Max-SINR
receiver is robust to the channel delay spread estimation errors.
\end{abstract}


%
\IEEEpeerreviewmaketitle

\section{Introduction}
\IEEEPARstart{O}{rthogonal} frequency division multiplexing (OFDM)
systems with guard-time interval or cyclic prefix can prevent
inter-symbol interference (ISI). OFDM has overlapping spectra and
rectangular impulse responses. Consequently, each OFDM sub-channel
exhibits a sinc-shape frequency response. Therefore, the time
variations of the channel during one OFDM symbol duration destroy
the orthogonality of different subcarriers, and result in power
leakage among subcarriers, known as inter-carrier interference
(ICI), which causes degradation in system performance. In order to
overcome the above drawbacks of OFDM system, several pulse-shaping
OFDM systems such as multiwavelets based OFDM system, OFDM based on
offset quadrature amplitude modulation (OFDM/OQAM) system, et al.,
were proposed
\cite{Kumb07,Das07,Abb10,GaoX11,Jun07,Ma08,Lin08,Sio02}.

It is shown in \cite{Str03,Han07,Han09,Han10} that signal
transmission through a rectangular lattice is suboptimal for doubly
dispersive (DD) channel. By using results from sphere covering
theory\cite{Con98}, the authors have demonstrated that lattice OFDM
(LOFDM) system, which is OFDM system based on hexagonal-type
lattice, providing better performance against ISI/ICI \cite{Str03}.
However, LOFDM confines the transmission pulses to a set of
orthogonal ones. In \cite{Han07,Han09,Han10}, the authors abandoned
the orthogonality condition of the modulated pulses and proposed a
multicarrier transmission scheme named as hexagonal multicarrier
transmission (HMT). In HMT system, there is no cyclic prefix and
data symbols of HMT signal are transmitted at the hexagonal type
lattice points in TF plane. In our previous work
\cite{Xu09,Xu11,Xu12,Xu12G}, we have analyzed the system SINR
performance and presented that traditional HMT receiver proposed in
\cite{Han07,Han09,Han10} using zero timing offset prototype pulse is
a suboptimal approach in the view of SINR. The receiver prototype
pulse based on Max-SINR criterion for HMT system over DD channel
with exponential power delay profile and U-shape Doppler spectrum
was proposed.

In this paper, we will present the receiver prototype pulses based
on Max-SINR criterion for HMT system over DD channel with different
channel scattering functions. Theoretical analyses show that there
is a timing offset between prototype pulses of the proposed Max-SINR
HMT receiver and traditional HMT receiver in the SINR point of view.
The closed form timing offset expressions of prototype pulse for
Max-SINR HMT receiver over the DD channel with different channel
scattering functions are derived. Simulation results show that the
proposed Max-SINR receiver obtains an approximation to the
theoretical upper bound SINR performance and outperforms traditional
receiver \cite{Han07,Han09,Han10} on bit error rate (BER)
performance. Meanwhile, the SINR performance of the proposed scheme
is robust to the channel delay spread estimation errors.

\section{Hexagonal Multicarrier Transmission System}
\begin{figure}[!t] \centering
\includegraphics[width=3in]{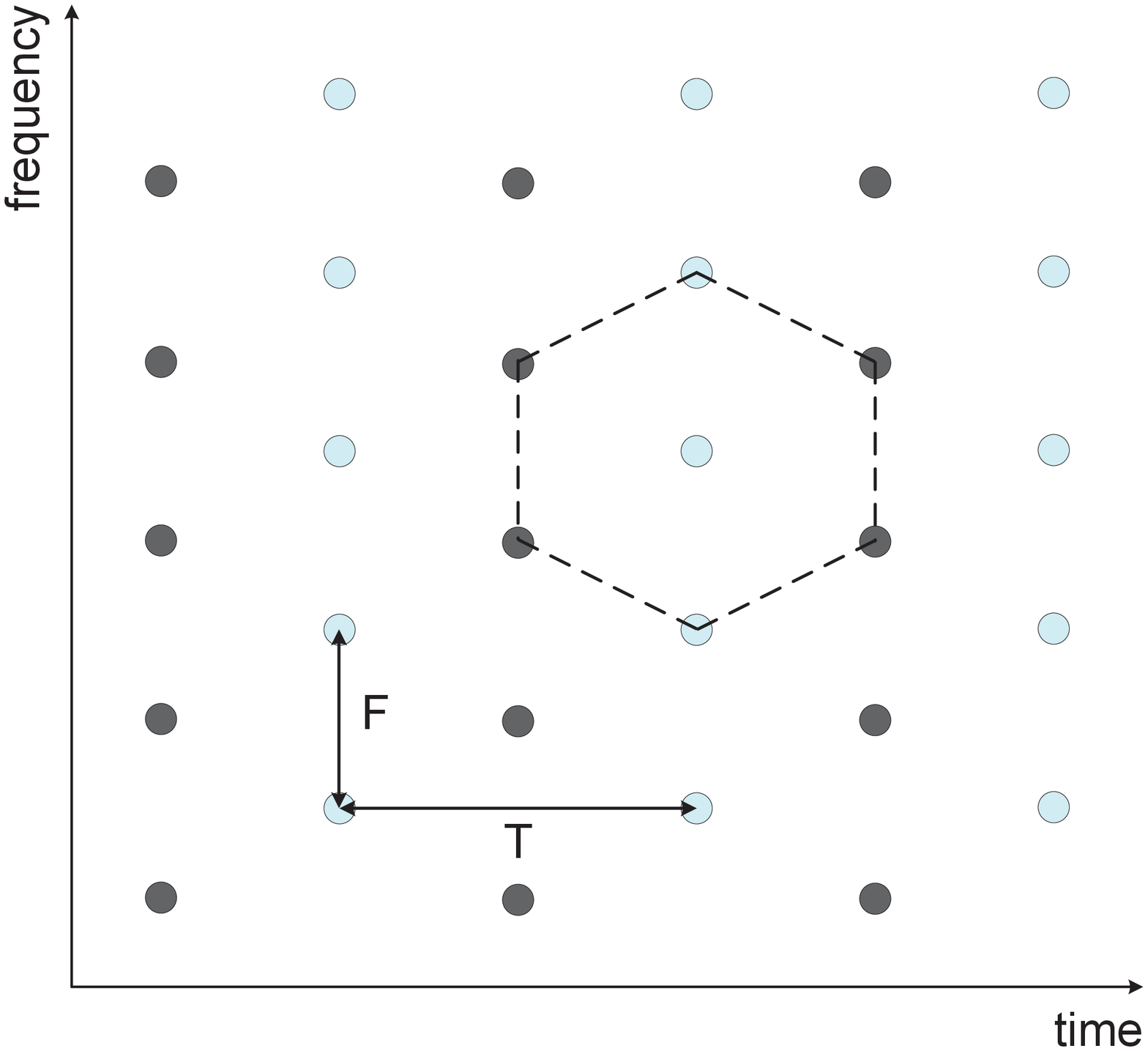}
\caption{Partition of the hexagonal lattice into a rectangular
sublattice $\textit{\textbf{V}}_{\textrm{rect}1}$ (denoted by
$\circ$) and its coset $\textit{\textbf{V}}_{\textrm{rect}2}$
(denoted by $\bullet$).} \label{fig_sim}
\end{figure}
In HMT systems, the transmitted baseband signal can be expressed as
\cite{Han07,Han09,Han10}
\begin{equation} \label{1}
\begin{split}
x(t)&=\sum_{m}\sum_{n}c_{m,2n}g\big(t-mT\big)e^{j2\pi
nFt}\\
&+\sum_{m}\sum_{n}c_{m,2n+1}g\big(t-mT-\frac{T}{2}\big)e^{j2\pi(nF+\frac{F}{2})t}
\end{split}
\end{equation}
where $T$ and $F$ are the lattice parameters, as shown in Fig.1;
$c_{m,n}$ denotes the transmitted data symbol, which is assumed to
be taken from a specific signal constellation and independent and
identically distributed (i.i.d.) with zero mean and average power
$\sigma_{c}^{2}$; $m\in\mathcal {M}$ and $n\in\mathcal {N}$ are the
position indices in the TF plane; $\mathcal {M}$ and $\mathcal {N}$
denote the sets from which $m,n$ can be taken, with cardinalities
$\textit{M}$ and $\textit{N}$, respectively. The prototype pulse
$g(t)$ is the Gaussian window\footnotemark \footnotetext{LOFDM
system confines the transmission pulses to a set of orthogonal ones.
As pointed out in \cite{Han07}, these orthogonalized pulses destroy
the TF concentration of the initial pulses, hence lower the
robustness to the time and frequency dispersion caused by the DD
channel. On the other hand, the orthogonalized pulses are
constructed from a nonorthogonal set of Gaussian pulses. As shown in
\cite{Str03}, the orthogonalization process needs to compute all the
elements of the Gram matrix, which leads to high implementation
complexity. } $g(t)=(2/\sigma)^{1/4}e^{-(\pi/\sigma)t^{2}}$. The
ambiguity function of Gaussian pulse is defined by
\begin{equation} \label{2}
\begin{split}
A_{g}(\tau,\upsilon)&=\int_{-\infty}^{\infty}g(t)g^{*}(t-\tau)e^{-j2\pi\upsilon
t}dt\\
&=e^{-\frac{\pi}{2}(\frac{1}{\sigma}\tau^{2}+\upsilon^{2})}e^{-j\pi\tau\upsilon}
\end{split}
\end{equation}
where $(\cdot)^{*}$ denotes the complex conjugate. The original
hexagonal lattice can be expressed as the disjoint union of a
rectangular sublattice $\textit{\textbf{V}}_{\textrm{rect}1}$ and
its coset $\textit{\textbf{V}}_{\textrm{rect}2}$, as shown in Fig.
1. The transmitted baseband signal in Eq. (1) can be rewritten as
$x(t)=\sum_{i}\sum_{m}\sum_{n}c_{m,n}^{i}g_{m,n}^{i}(t)$, where
$i$=$1,2$, $c_{m,n}^{1}$ and $c_{m,n}^{2}$ represent symbols coming
from $\textit{\textbf{V}}_{\textrm{rect}1}$ and
$\textit{\textbf{V}}_{\textrm{rect}2}$, respectively.
$g_{m,n}^{i}(t)=g(t-mT-\frac{(i-1)T}{2})e^{j2\pi(nF+\frac{(i-1)F}{2})t}$
is the transmitted pulse corresponding to $c_{m,n}^{i}$. The
baseband DD channel can be modeled as a random linear operator
$\textrm{H}$ \cite{Bel63}
\begin{equation} \label{3}
\textrm{H}[x(t)]=\int_{0}^{\tau_{\textrm{max}}}\int^{f_{d}}_{-f_{d}}H(\tau,\upsilon)x(t-\tau)e^{j2\pi\upsilon
t}d\tau d\upsilon
\end{equation}
where $\tau_{\textrm{max}}$ and $f_{d}$ are the maximum multipath
delay spread and the maximum Doppler frequency, respectively. The
product $\vartheta=\tau_{\textrm{max}}f_{d}$ is referred to as the
channel spread factor (CSF)\cite{Bel63,Coh95} and $H(\tau,\upsilon)$
is called the delay-Doppler spread function. In wide-sense
stationary uncorrelated scattering (WSSUS) assumption the DD channel
is characterized by the second-order statistics
\begin{equation} \label{4}
\textrm{E}[H(\tau,\upsilon)H^{*}(\tau_{1},\upsilon_{1})]=S_{H}(\tau,\upsilon)\delta(\tau-\tau_{1})\delta(\upsilon-\upsilon_{1})
\end{equation}
where $\textrm{E}[\cdot]$ denotes the expectation and
$S_{H}(\tau,\upsilon)$ is called the scattering function. Without
loss of generality, we use
$\int_{0}^{\tau_{\textrm{max}}}\int_{-f_{d}}^{f_{d}}S_{H}(\tau,\upsilon)d\tau
d\upsilon=1$, that is the channel has no overall path loss. The
received signal can be expressed as $r(t)=\textrm{H}[x(t)]+w(t)$,
where $w(t)$ is the AWGN with variance $\sigma_{w}^{2}$.

\section{The Max-SINR Receiver}
To obtain the data symbol $\hat{c}_{m,n}^{i}$, the match filter
receiver projects the received signal $r(t)$ on prototype pulse
function $\psi_{m,n}^{i}(t),i=1,2$, i.e.,
\begin{equation} \label{5}
\begin{split}
\hat{c}_{m,n}^{i}&=\big<r(t),\psi_{m,n}^{i}(t)\big>
\\
&=\sum_{j}\sum_{m',n'}c_{m',n'}^{j}\big<\textrm{H}[g_{m',n'}^{j}(t)],\psi_{m,n}^{i}(t)\big>\\
&+\big<w(t),\psi_{m,n}^{i}(t)\big>
\end{split}
\end{equation}
where
$\psi_{m,n}^{i}(t)$=$\psi(t-mT-\frac{(i-1)T}{2})e^{j2\pi(nF+\frac{(i-1)F}{2})t}$,
and $\psi(t)$ is the prototype pulse at the receiver. The energy of
the received symbol $c_{m,n}^i$, after projection on the filter
function $\psi_{m,n}^{i}(t)$ over DD channel can be expressed as
\begin{equation} \label{6}
\begin{split}
E_{s}&=\textrm{E}\Bigg\{\Big|\sum_j\sum_{m',n'}c_{m',n'}^{j}\big<\textrm{H}[g_{m',n'}^{j}(t)],\psi_{m,n}^{i}(t)\big>\\
&+\big<w(t),\psi_{m,n}^{i}(t)\big>\Big|^2\Bigg\}
\end{split}
\end{equation}
Under the assumptions that WSSUS channel and source symbols
$c_{m,n}^{i}$ are statistically independent, (6) can be rewritten as
\begin{equation} \label{7}
\begin{split}
E_{s}&=\sigma_{c}^{2}\int_{\tau}\int_{\upsilon}S_{H}(\tau,\upsilon)\bigg[\sum_{m,n}\big|A_{g,\psi}(mT+\tau, nF+\upsilon)\big|^{2}\\
&+\sum_{m,n}\big|A_{g,\psi}(mT+\frac{T}{2}+\tau,
nF+\frac{F}{2}+\upsilon)\big|^{2}\bigg]d\tau
 d\upsilon\\
 &+\sigma_{w}^{2}\big|A_{g,\psi}(0,0)\big|
\end{split}
\end{equation}
We can see from Eq. (7) that $E_{s}$ is composed of the expectation
symbol energy, ISI/ICI and additive noise. The SINR of the desired
symbol $c_{m_0,n_0}^{i_0}$ can be expressed as
\begin{equation} \label{8}
R_{\textrm{SIN}}=\frac{\sigma_{c}^{2}}{E_{\textrm{IN}}}\int_{\tau}\int_{\upsilon}S_{H}(\tau,\upsilon)\big|A_{g,\psi}(\tau,\upsilon)\big|^2d\tau
d\upsilon
\end{equation}
where the interference-plus-noise energy $E_{\textrm{IN}}$ is the
energy perturbation of the desired symbol $c_{m_0,n_0}^{i_0}$, from
other symbols over time varying multipath fading channel
$H(\tau,\upsilon)$. For presentation simplicity, the desired symbol
$c_{m_0,n_0}^{i_0}$ is chosen as $c_{0,0}^{1}$. Hence,
$E_{\textrm{IN}}$ can be expressed as
\begin{equation} \label{9}
\begin{split}
E_{\textrm{IN}}&=\sigma_{c}^{2}\int_{\tau}\int_{\upsilon}S_{H}(\tau,\upsilon)\\
&\cdot\bigg[\sum_{[m,n,i]\neq[0,0,1]}\big|A_{g,\psi}\big((m+\frac{i}{2})T+\tau, nF+\upsilon\big)\big|^{2}\\
&+\sum_{[m,n,i]\neq[0,0,1]}\big|A_{g,\psi}\big((m+\frac{i}{2})T+\frac{T}{2}+\tau,
nF\\
&+\frac{F}{2}+\upsilon\big)\big|^{2}\bigg]d\tau
 d\upsilon+\sigma_{w}^{2}\big|A_{g,\psi}(0,0)\big|
\end{split}
\end{equation}
According to the form of channel scattering functions, we have the
following two cases \cite{Han07}.

\subsection{DD channel with uniform power delay profile and uniform Doppler
spectrum} For the DD channel with uniform power delay profile and
uniform Doppler spectrum (DD-UNI), the scattering function can be
expressed as $S_{H}(\tau,\upsilon)=1/(2\tau_{\textrm{max}}f_{d})$
\cite{Mat02}, with $\tau_{max}\geq\tau>0,|\upsilon|<f_{d}$. We
assume that\footnotemark \footnotetext{It is shown in
\cite{Wu05,Wu07,Wu071} that the optimum sampling time of wireless
communication systems over DD channel is dependent on the power
distribution of the channel profiles, and that zero timing offset
does not always yield the best system performance. Moreover, it is
shown in \cite{Das07} that there is a delay between the transmitted
Gaussian prototype pulse and the Max-SINR prototype pulse of
multicarrier transmission system with rectangular TF lattice over DD
channel. Meanwhile, the DD propagation channel causes transmission
signal dispersion in both the time and frequency domains. Hence, we
assume that there is a timing offset and a frequency offset between
the prototype pulses at the transmitter and the receiver.}
$\psi(t)=g(t-\Delta t)e^{j2\pi\Delta ft}$, $\mid\Delta f\mid<f_{d}$,
the SINR of the received signal over the DD-UNI channel can be
expressed as
\begin{equation} \label{10}
\begin{split}
R_{\textrm{SIN}}^{\textrm{UNI}}&=\frac{\sigma_{c}^{2}}{2\tau_{\textrm{max}}f_{d}E_{\textrm{IN}}^{\textrm{UNI}}}\int_{0}^{\tau_{\textrm{max}}}e^{-\frac{\pi}{\sigma}(\tau-\Delta
t)^{2}}d\tau\\
&\cdot\int_{-f_{d}}^{f_{d}}e^{-\sigma\pi(\upsilon-\Delta
f)^{2}}d\upsilon
\end{split}
\end{equation}
The theoretical SINR upper bound of the received signal over the
DD-UNI channel can be expressed as
\begin{equation}\label{11}
R_{\textrm{UB}}^{\textrm{UNI}}=\arg\max_{\Delta t, \Delta
f}R_{\textrm{SIN}}^{\textrm{UNI}}
\end{equation}
The Max-SINR prototype pulse can be expressed as (see Appendix A)
\begin{equation} \label{12}
\psi(t)=g(t-\frac{\tau_{\textrm{max}}}{2})
\end{equation}

\subsection{DD channel with exponential power delay profile and
U-shape Doppler spectrum} For the DD channel with exponential power
delay profile and U-shape Doppler spectrum (DD-EXP), the scattering
function can be expressed as \cite{Mat02}
\begin{equation} \label{13}
S_{H}(\tau,\upsilon)=\frac{e^{\frac{-\tau}{\tau_{\textrm{rms}}}}}{\pi\tau_{\textrm{rms}}f_{d}\sqrt{1-(\upsilon/f_{d})^2}}
\end{equation}
with $\tau>0,|\upsilon|<f_{d}$, $\tau_{\textrm{rms}}$ denotes the
channel root mean square (RMS) delay spread. We assume that
$\psi(t)=g(t-\Delta t)e^{j2\pi\Delta ft}$, $\mid\Delta f\mid<f_{d}$,
the theoretical SINR can be expressed as
\begin{equation} \label{14}
\begin{split}
R_{\textrm{SIN}}^{\textrm{EXP}}&=\frac{\sigma_{c}^{2}}{\pi\tau_{\textrm{rms}}f_{d}E_{\textrm{IN}}^{\textrm{EXP}}}\int_{0}^{\infty}e^{-\frac{\tau}{\tau_{\textrm{rms}}}}e^{-\frac{\pi}{\sigma}(\tau-\Delta
t)^{2}}d\tau\\
&\cdot\int_{-f_{d}}^{f_{d}}\frac{e^{-\sigma\pi(\upsilon-\Delta
f)^{2}}}{\sqrt{1-(\upsilon/f_{d})^{2}}}d\upsilon
\end{split}
\end{equation}
The theoretical SINR upper bound of the received signal over the
DD-EXP channel can be expressed as
\begin{equation}\label{15}
R_{\textrm{UB}}^{\textrm{EXP}}=\arg\max_{\Delta t, \Delta
f}R_{\textrm{SIN}}^{\textrm{EXP}}
\end{equation}
Plugging (13) in (15), the Max-SINR prototype pulse can be expressed
as $\psi(t)=g(t-\Delta t)$ and (see Appendix B)\footnotemark
\footnotetext{In \cite{Das07}, the Max-SINR prototype pulse
$\textit{\textbf{g}}$ of multicarrier transmission system with
rectangular TF lattice over DD channel is obtained by maximizing the
generalized Rayleigh quotient
$\hat{\textit{\textbf{g}}}=arg\max\frac{\textit{\textbf{g}}^H\textit{\textbf{B}}\textit{\textbf{g}}}{\textit{\textbf{g}}^H\textit{\textbf{A}}\textit{\textbf{g}}}$.
The solution is the generalized eigenvector of the matrix pair
$(\textit{\textbf{B}},\textit{\textbf{A}})$ corresponding to the
largest generalized eigenvalue. It is shown that there is a delay
between the transmitted Gaussian prototype pulse and the received
Gaussian prototype pulse. In this paper, the close form time offset
expressions between the transmitted and received prototype pulse of
multicarrier transmission system with hexagonal TF lattice is
derived.}
\begin{equation} \label{16}
\begin{split}
\Delta
t&=\frac{\sigma}{2\pi\tau_{\textrm{rms}}}-\sqrt{\frac{\sigma}{2\pi}}\Bigg(\frac{\frac{3.28\sqrt{\sigma}}{\tau_{\textrm{rms}}}-\sqrt{\frac{3.28^2\sigma}{\tau_{\textrm{rms}}^2}-3.52\big(\frac{\sigma}{\tau^2_{\textrm{rms}}}-4}\big)}{1.76}\Bigg)
\end{split}
\end{equation}

We can see from equation (12) and (16) that the prototype pulses of
the proposed Max-SINR receiver over DD channel are functions of
channel maximum delay spread and RMS delay spread, respectively.
Recently, several channel estimation schemes for multicarrier
modulation system with hexagonal TF lattice have been proposed in
\cite{Gao09,Gao11,Gao11F} and all this schemes are suitable for HMT
system.

\section{Simulation and Discussion}
In this section, we test the proposed Max-SINR receiver via computer
simulations based on the discrete signal model. In the following
simulations, the number of subcarriers for HMT system is chosen as
$N$=40, and the length of prototype pulse is set to $N_{g}$=600. The
center carrier frequency is $f_{c}$=5GHz and the sampling interval
is set to $T_{s}$=$10^{-6}s$. The system parameters of HMT system
are $F$=25kHz, $T$=$1\times10^{-4}s$ and the signaling efficiency
$\rho$=0.8. $\sigma$ for prototype pulse $g(t)$ is set to
$\sigma$=$T/\sqrt{3}F$. Traditional projection receiver proposed in
\cite{Han07,Han09,Han10} is named as Traditional Projection Receiver
(TPR) in the following simulation results.

\subsection{SINR Performance of the Proposed Max-SINR Receiver for HMT System}
\subsubsection{SINR Performance of HMT System over DD-UNI Channel} The SINR
performance of different receivers with the variety of
$\sigma_{c}^{2}/\sigma_{w}^{2}$ for HMT system over DD-UNI channel
is depicted in Fig. 2. The CSF $\vartheta$ is set to 0.07 and 0.2,
respectively. We can see from Fig. 2 that the SINR performance of
Max-SINR receiver outperforms TPR scheme about
0.5$\sim1.5\textrm{dB}$ at $\vartheta$=0.07 and 1$\sim3\textrm{dB}$
at $\vartheta=0.2$, respectively. The SINR gap between Max-SINR
receiver and theoretical SINR upper bound is smaller than 0.1dB at
$\vartheta=0.07$ and 0.2, respectively.
\begin{figure}[!t]
\centering
\includegraphics[width=3.3in]{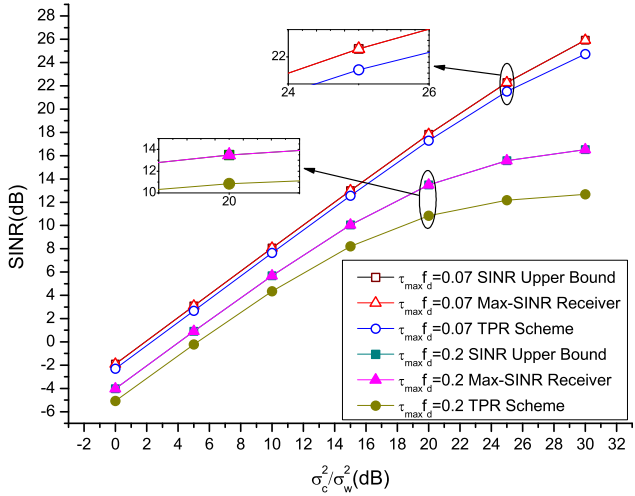}
\caption{The SINR performance of different receivers with the
variety of $\sigma_{c}^2/\sigma_{w}^2$ for HMT system over DD-UNI
channel.} \label{fig_sim}
\end{figure}
The SINR performance with the variety of channel spread factors
$\vartheta$ at $\sigma_{c}^2/\sigma_{w}^2$=20dB over DD-UNI channel
is depicted in Fig. 3. It can be seen that there is a degradation of
SINR with the increasing of channel spread factor. Max-SINR receiver
obtains an approximation to the theoretical upper bound SINR
performance within the full range of $\vartheta$. Meanwhile,
Max-SINR receiver obtains an about 3.5dB maximum SINR gain over TPR
scheme at $\vartheta$=0.35.
\begin{figure}[!t]
\centering
\includegraphics[width=3.3in]{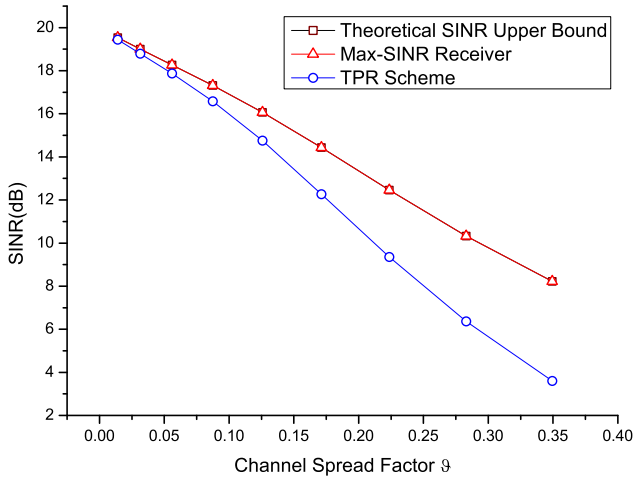}
\caption{The SINR performance of different receivers with the
variety of channel spread factors $\vartheta$ for HMT system over
DD-UNI channel, $\sigma_{c}^2/\sigma_{w}^2$=20dB. } \label{fig_sim}
\end{figure}

\subsubsection{Performance of HMT System over DD-EXP Channel} The
SINR performance of different receivers with the variety of
$\sigma_{c}^{2}/\sigma_{w}^{2}$ for HMT system over DD-EXP channel
is depicted in Fig. 4. $\tau_{\textrm{rms}}f_d$ in Fig. 4 is set to
0.07 and 0.2, respectively. We can see from Fig. 4 that the SINR
performance of Max-SINR receiver outperforms TPR scheme about
1$\sim4\textrm{dB}$ at $\tau_{\textrm{rms}}f_d$=0.07 and
1.5$\sim3.5\textrm{dB}$ at $\tau_{\textrm{rms}}f_d=0.2$,
respectively. The SINR gap between Max-SINR receiver and the
theoretical SINR upper bound is smaller than 0.5dB and 0.1dB at
$\tau_{\textrm{rms}}f_d=0.07$ and 0.2, respectively.
\begin{figure}[!t]
\centering
\includegraphics[width=3.3in]{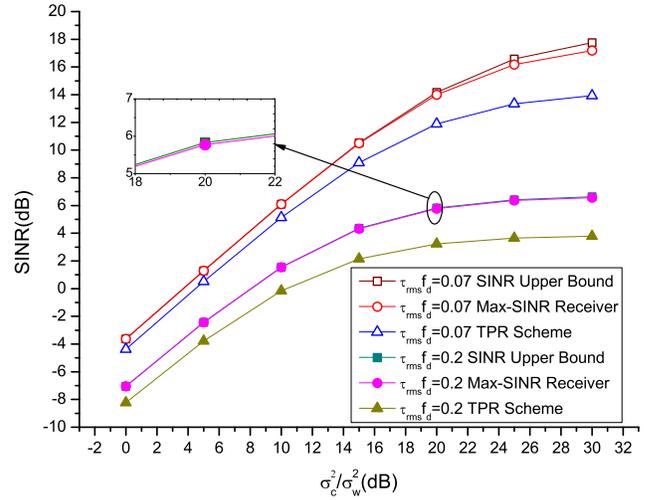}
\caption{The SINR performance of different receivers with the
variety of $\sigma_{c}^2/\sigma_{w}^2$ for HMT system over DD-EXP
channel.} \label{fig_sim}
\end{figure}
The SINR performance with the variety of $\tau_{\textrm{rms}}f_d$ at
$\sigma_{c}^2/\sigma_{w}^2$=20dB over DD-EXP channel is depicted in
Fig. 5. It can be seen that there is a degradation of SINR with the
increasing of $\tau_{\textrm{rms}}f_d$. Max-SINR receiver obtains an
approximation to the theoretical upper bound SINR performance within
the full range of $\tau_{\textrm{rms}}f_d$. Max-SINR receiver
achieves an about 2.5dB maximum SINR gain over TPR scheme at
$\tau_{\textrm{rms}}f_d$=0.35.
\begin{figure}[!t]
\centering
\includegraphics[width=3.3in]{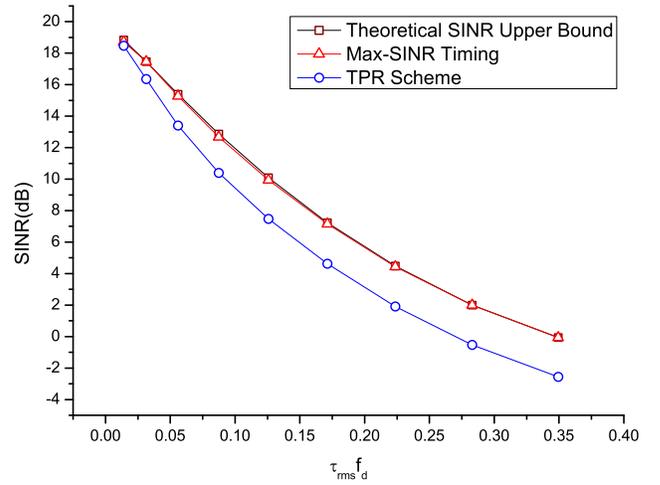}
\caption{The SINR performance of different receivers with the
variety of $\tau_{\textrm{rms}}f_d$ for HMT system over DD-EXP
channel, $\sigma_{c}^2/\sigma_{w}^2$=20dB.} \label{fig_sim}
\end{figure}

\subsection{BER Performance of the Proposed Max-SINR Receiver for HMT System}
The BER performance of the proposed Max-SINR receiver for HMT system
over the DD channel with different channel scattering functions is
given in Fig. 6.

For the DD-UNI channel, channel spread factor
$\tau_{\textrm{max}}f_d$ is set to 0.2. We can conclude from Fig. 6
that the BER performance of the proposed Max-SINR receiver for HMT
system over DD-UNI channel outperforms TPR receiver about 2dB at
$E_b/N_0=20$dB. For the DD-EXP channel, $\tau_{\textrm{rms}}f_d$ is
set to 0.1. We can see from Fig. 6 that the proposed Max-SINR
receiver over DD-EXP channel outperforms TPR receiver on the BER
performance and the performance gain is about 2.5dB at
$E_b/N_0=20$dB.

\begin{figure}[!t]
\centering
\includegraphics[width=3.3in]{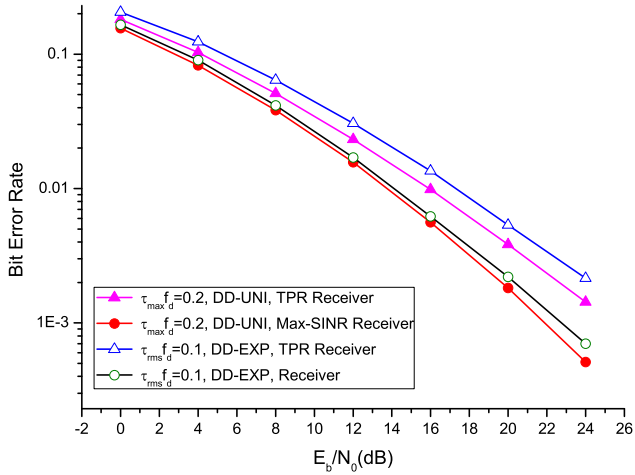}
\caption{The BER performance of the proposed Max-SINR receiver for
HMT system over the DD channel with different channel scattering
functions.} \label{fig_sim}
\end{figure}

\subsection{Robustness of the Proposed Max-SINR Receiver against Channel Delay Spread Estimation Errors}
The robustness of the proposed Max-SINR receiver against channel
delay spread estimation errors is depicted in Fig. 7. Estimation
errors of $\tau_{\textrm{rms}}$ and $\tau_{\textrm{max}}$ for DD-EXP
channel and DD-UNI channel are modeled as uniformly distributed
random variables in the interval
$[-\tau_{\textrm{rms}}/2,\tau_{\textrm{rms}}/2]$ and
$[-\tau_{\textrm{max}}/2,\tau_{\textrm{max}}/2]$, respectively.

\begin{figure}[!t]
\centering
\includegraphics[width=3.3in]{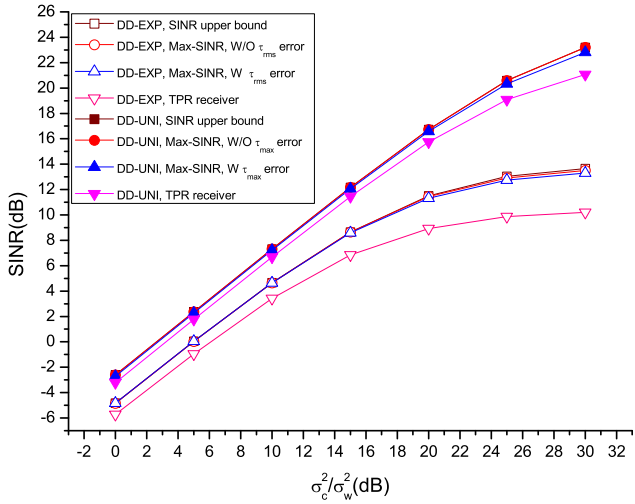}
\caption{The robustness of the proposed Max-SINR receiver against
channel delay spread estimation errors. Estimation errors of
$\tau_{\textrm{rms}}$ and $\tau_{\textrm{max}}$ for DD-EXP channel
and DD-UNI channel are modeled as uniformly distributed random
variables in the interval
$[-\tau_{\textrm{rms}}/2,\tau_{\textrm{rms}}/2]$ and
$[-\tau_{\textrm{max}}/2,\tau_{\textrm{max}}/2]$, respectively.}
\label{fig_sim}
\end{figure}

In Fig.7, the channel spread factor $\vartheta$ of DD-UNI channel is
set to 0.1. We can see from Fig. 7 that the SINR gap between the
SINR upper bound and the proposed Max-SINR receiver with estimation
error of $\tau_{\textrm{max}}$ is within 0.1dB and 0.5dB at
$\sigma_{c}^2/\sigma_{w}^2$=0dB and 30dB, respectively. The SINR
performance of the proposed Max-SINR receiver over DD-EXP channel at
$\tau_{\textrm{rms}}f_d$=0.1 is given in Fig. 7. The SINR gap
between the SINR upper bound and the proposed Max-SINR receiver with
estimation error of $\tau_{\textrm{rms}}$ is within 0.1dB and 0.7dB
at $\sigma_{c}^2/\sigma_{w}^2$=0dB and 30dB, respectively. SINR
performance of the TPR receiver over DD-UNI and DD-EXP channel is
also depicted in Fig. 7 for comparison. The proposed Max-SINR
receiver outperforms TPR scheme when there exists estimation errors
of $\tau_{\textrm{rms}}$ and $\tau_{\textrm{max}}$.

\section{Conclusion}
A novel receiver based on Max-SINR criterion for HMT system over DD
channel with different channel scattering functions is proposed in
this paper. Theoretical analyses show that there is a timing offset
between the prototype pulses of the proposed Max-SINR receiver and
the traditional projection receiver. The closed form timing offset
expressions of prototype pulse for Max-SINR HMT receiver over DD
channel with different channel scattering functions are derived.
Simulation results show that the proposed Max-SINR receiver obtains
an approximation to the theoretical upper bound SINR performance and
outperforms traditional projection scheme on BER performance.
Meanwhile, the SINR performance of the proposed prototype pulse is
robust to the channel delay spread estimation errors.

\appendices
\section{Proof of Equation (12)}
The SINR of the received signal over DD-UNI channel can be expressed
as
\begin{equation} \label{17}
R_{\textrm{SIN}}^{\textrm{UNI}}\approx\underbrace{\frac{\sigma_{c}^{2}}{2\sigma_{w}^{2}\tau_{\textrm{max}}f_{d}}\int_{0}^{\tau_{\textrm{max}}}e^{\frac{-\pi(\tau-\Delta
t)^2}{\sigma}}d\tau}_{R^{\textrm{UNI}}(\Delta
t)}\underbrace{\int_{-f_{d}}^{f_{d}}e^{-\sigma\pi(\upsilon-\Delta
f)^{2}}d\upsilon}_{R^{\textrm{UNI}}(\Delta f)}
\end{equation}
We can see from equation (17) that $R_{\textrm{SIN}}^{\textrm{UNI}}$
is the product of two functions $R^{\textrm{UNI}}(\Delta t)$ and
$R^{\textrm{UNI}}(\Delta f)$ with respect to $\Delta t$ and $\Delta
f$, respectively. Hence, the optimal timing offset $\Delta t$ and
the optimal frequency offset $\Delta f$ can be obtained
independently.

The optimal timing offset $\Delta t$ can be obtained by taking the
partial derivative of $R_{\textrm{SIN}}^{\textrm{UNI}}$ with respect
to ${\Delta t}$ and solving the partial derivative equal to zero for
${\Delta t}$,
\begin{equation} \label{18}
\frac{\partial R_{\textrm{SIN}}^{\textrm{UNI}}}{\partial \Delta t}=0
\end{equation}
Plugging (17) in (18)  and ignoring the constant items with respect
to $\Delta t$, the partial derivative can be rewritten as
\begin{equation} \label{19}
\int_{0}^{\tau_{\textrm{max}}}\frac{2\pi(\tau-\Delta
t)}{\sigma}e^{\frac{-\pi(\tau-\Delta t)^2}{\sigma}}d\tau=0
\end{equation}
Under the assumption that $0<\Delta t<\tau_{max}$, we can rewrite
(19) as
\begin{equation} \label{20}
\begin{split}
\int_{\Delta t}^{\tau_{max}}\frac{2\pi(\tau-\Delta
t)}{\sigma}e^{\frac{-\pi(\tau-\Delta
t)^2}{\sigma}}d\tau\\
+\int_{0}^{\Delta t}\frac{2\pi(\tau-\Delta
t)}{\sigma}e^{\frac{-\pi(\tau-\Delta t)^2}{\sigma}}d\tau=0
\end{split}
\end{equation}
Let $\tau'=\tau-\Delta t$, we can rewrite (20) as
\begin{equation} \label{21}
\begin{split}
\underbrace{\int_{0}^{\tau_{max}-\Delta
t}\frac{2\pi\tau'}{\sigma}e^{\frac{-\pi(\tau')^2}{\sigma}}d\tau'}_{\alpha(\Delta
t)}=\underbrace{\int_{0}^{\Delta
t}\frac{2\pi\tau'}{\sigma}e^{\frac{-\pi(\tau')^2}{\sigma}}d\tau'}_{\beta(\Delta
t)}
\end{split}
\end{equation}
Hence, the solution of equation $\partial
R_{\textrm{SIN}}^{\textrm{UNI}}/\partial \Delta t=0$ is
$\tau_{max}-\Delta t = \Delta t$, that is $\Delta
t=\tau_{\textrm{max}}/2$, and the SINR of the received symbols
obtains the maximum value while $\Delta t=\tau_{\textrm{max}}/2$.

The optimal frequency offset $\Delta f$ can be obtained by taking
the partial derivative of $R_{\textrm{SIN}}^{\textrm{UNI}}$ with
respect to ${\Delta f}$ and solving the partial derivative equal to
zero for ${\Delta f}$,
\begin{equation} \label{22}
\frac{\partial R_{\textrm{SIN}}^{\textrm{UNI}}}{\partial \Delta f}=0
\end{equation}
Plugging (17) in (22) and let $\upsilon'=\upsilon-\Delta f$, we can
rewrite (22) as
\begin{equation} \label{23}
\begin{split}
\underbrace{\int_{0}^{f_{d}-\Delta
f}2\sigma\pi\upsilon'e^{-\sigma\pi(\upsilon')^{2}}d\upsilon'}_{\kappa
(\Delta f)}=\underbrace{\int_{0}^{f_{d}+\Delta
f}2\sigma\pi\upsilon'e^{-\sigma\pi(\upsilon')^{2}}d\upsilon'}_{\chi
(\Delta f)}
\end{split}
\end{equation}
Hence, the solution of partial derivative $\partial
R_{\textrm{SIN}}^{\textrm{UNI}}/\partial \Delta f=0$ is $f_d-\Delta
f=f_d+\Delta f$, that is $\Delta f=0$.

\section{Proof of Equation(16)}
The SINR of the received signal can be expressed as
\begin{equation} \label{24}
\begin{split}
R_{\textrm{SIN}}^{\textrm{EXP}}&\approx\frac{\sigma_{c}^{2}}{\sigma_{w}^{2}\tau_{\textrm{rms}}f_{d}}\int_{-f_{d}}^{f_{d}}\frac{e^{-\sigma\pi(\upsilon-\Delta
f)^{2}}}{\sqrt{1-(\upsilon/f_{d})^2}}d\upsilon\\
&\cdot\Bigg(\underbrace{e^{\frac{\sigma}{4\pi\tau_{\textrm{rms}}^{2}}+\frac{\Delta
t}{\tau_{\textrm{rms}}}}}_{a(\Delta
t)}\underbrace{\int_{0}^{\infty}e^{-\frac{\pi}{\sigma}(\tau-\Delta
t+\frac{\sigma}{2\pi\tau_{\textrm{rms}}})^2}d\tau}_{b(\Delta
t)}\Bigg)
\end{split}
\end{equation}
where
\begin{equation} \label{25}
\begin{split}
b(\Delta
t)&=\sqrt{\frac{\sigma}{\pi}}\int^{\infty}_{\sqrt{\frac{\pi}{\sigma}}(\frac{\sigma}{2\pi\tau_{\textrm{rms}}}-\Delta
t)}e^{-x^2}dx\\
&=\frac{\sqrt{\sigma}}{2}\textrm{erfc}\Bigg(\sqrt{\frac{\pi}{\sigma}}\Big(\frac{\sigma}{2\pi\tau_{\textrm{rms}}}-\Delta
t\Big)\Bigg)
\end{split}
\end{equation}
where $\textrm{erfc}(\cdot)$ is the complementary error function. If
$x>0$, we may obtain an approximate solution of the complementary
error function $\textrm{erfc}(\cdot)$ by \cite{Kin05}
\begin{equation} \label{26}
\textrm{erfc}(\frac{x}{\sqrt{2}})\simeq\frac{2e^{-\frac{x^2}{2}}}{1.64x+\sqrt{0.76x^2+4}}
\end{equation}
We can see from equation (24) that $R_{\textrm{SIN}}^{\textrm{UNI}}$
is also the product of two functions with respect to $\Delta t$ and
$\Delta f$, respectively. Hence, the optimal timing offset $\Delta
t$ and the optimal frequency offset $\Delta f$ can be obtained
independently.

The optimal timing offset $\Delta t$ can be obtained by solving the
gradient $a(\Delta t)b(\Delta t)$ with respect to $\Delta t$ to
zero,
\begin{equation} \label{27}
\frac{d a(\Delta t)}{d \Delta t}b(\Delta t)+\frac{d b(\Delta t)}{d
\Delta t}a(\Delta t)=0
\end{equation}
where $\frac{d a(\Delta t)}{d \Delta t}=-\frac{a(\Delta
t)}{\tau_{\textrm{rms}}}$ and $\frac{d b(\Delta t)}{d \Delta
t}=e^{-\frac{\pi}{\sigma}\big(\frac{\sigma}{2\pi\tau_{\textrm{rms}}}-\Delta
t\big)^2}$. Hence, equation (27) can be rewritten as
\begin{equation}
\begin{split} \label{28}
\frac{b(\Delta
t)}{\tau_{\textrm{rms}}}&=e^{-\frac{\pi}{\sigma}\big(\frac{\sigma}{2\pi\tau_{\textrm{rms}}}-\Delta
t\big)^2}\\
&=\frac{\sqrt{\sigma}}{2\tau_{\textrm{rms}}}\textrm{erfc}\bigg(\sqrt{\frac{\pi}{\sigma}}\big(\frac{\sigma}{2\pi\tau_{\textrm{rms}}}-\Delta
t\big)\bigg)\\
&\simeq\frac{\sqrt{\sigma}}{\tau_{\textrm{rms}}}e^{-\frac{\pi}{\sigma}\big(\frac{\sigma}{2\pi\tau_{\textrm{rms}}}-\Delta
t\big)^2}\Bigg(1.64\sqrt{\frac{2\pi}{\sigma}}\big(\frac{\sigma}{2\pi\tau_{\textrm{rms}}}-\Delta
t\big)\\
&+\sqrt{\frac{1.52\pi}{\sigma}\big(\frac{\sigma}{2\pi\tau_{\textrm{rms}}}-\Delta
t\big)^2+4}\Bigg)^{-1}
\end{split}
\end{equation}
Equation (28) can be simplified to a quadratic equation. Under the
constraint of $\Delta t>0$, the solution of the quadratic equation
can be expressed as
\begin{equation} \label{29}
\begin{split}
\Delta
t&=\frac{\sigma}{2\pi\tau_{\textrm{rms}}}-\sqrt{\frac{\sigma}{2\pi}}\Bigg(\frac{\frac{3.28\sqrt{\sigma}}{\tau_{\textrm{rms}}}-\sqrt{\frac{3.28^2\sigma}{\tau_{\textrm{rms}}^2}-3.52\big(\frac{\sigma}{\tau^2_{\textrm{rms}}}-4}\big)}{1.76}\Bigg)
\end{split}
\end{equation}

The optimal timing offset $\Delta f$ can be obtained by solving the
partial derivative $R_{\textrm{SIN}}^{\textrm{EXP}}$ with respect to
$\Delta f$ to zero,
\begin{equation} \label{30}
\frac{\partial R_{\textrm{SIN}}^{\textrm{EXP}}}{\partial \Delta f}=0
\end{equation}
Let $\Xi(\Delta f)$ denotes the partial derivative $\partial
R_{\textrm{SIN}}^{\textrm{EXP}}/{\partial \Delta f}$. Plugging (24)
in (30) and ignoring the constant items with respect to $\Delta f$,
$\Xi(\Delta f)$ can be rewritten as
\begin{equation} \label{31}
\begin{split}
\Xi(\Delta f)&=\int_{-f_{d}}^{f_{d}}\frac{2\sigma\pi(\upsilon-\Delta
f)e^{-\sigma\pi(\upsilon-\Delta
f)^{2}}}{\sqrt{1-(\upsilon/f_{d})^{2}}}d\upsilon
\end{split}
\end{equation}
Both of the exponential function and $\sqrt{1-(\upsilon/f_{d})^{2}}$
are non-negative, and $2\sigma\pi(\upsilon-\Delta f)$ is a
monotonically increasing function. $\Xi(-\Delta f)$ can be expressed
as
\begin{equation} \label{32}
\begin{split}
\Xi(-\Delta
f)&=\int_{-f_{d}}^{f_{d}}\frac{2\sigma\pi(\upsilon+\Delta
f)e^{-\sigma\pi(\upsilon+\Delta
f)^{2}}}{\sqrt{1-(\upsilon/f_{d})^{2}}}d\upsilon\\
&=-\Xi(\Delta f)
\end{split}
\end{equation}
Hence, $\Xi(\Delta f)$, $|\Delta f|<f_{d}$, is a continuous odd
function. Meanwhile,
$\partial^2R_{\textrm{SIN}}^{\textrm{EXP}}/{\partial\Delta
f^2}=\partial\Xi(\Delta f)/\partial\Delta f$ can be expressed as
\begin{equation} \label{33}
\begin{split}
\frac{\partial\Xi(\Delta f)}{\partial\Delta
f}&=\int_{-f_{d}}^{f_{d}}
\frac{2\sigma\pi\big(2\sigma\pi(\upsilon-\Delta
f)^2-1\big)e^{-\sigma\pi(\upsilon-\Delta
f)^{2}}}{\sqrt{1-(\upsilon/f_{d})^{2}}}d\upsilon
\end{split}
\end{equation}
Practical wireless channels usually satisfy that
$\vartheta=\tau_{\textrm{max}}f_{d}\ll1$\cite{Bel63} and the optimal
system parameter for HMT over DD channels can be chosen as
$\sigma=\tau_{\textrm{max}}/f_{d}$ \cite{Han07}. Hence, $\sigma$ in
(33) satisfies $\sigma\ll 1/\Delta f^2$ and $\partial\Xi(\Delta
f)/\partial\Delta f<0$, $|\Delta f|<f_{d}$. We can conclude from
equation (32) and (33) that $\Xi(\Delta f)$, $|\Delta f|<f_{d}$, is
a continuous odd function and $\partial\Xi(\Delta f)/\partial\Delta
f<0$, therefore, the necessary and sufficient condition of
$R_{\textrm{SIN}}^{\textrm{EXP}}$ to obtain its maximum value is
$\Delta f=0$.

\ifCLASSOPTIONcaptionsoff
  \newpage
\fi

\end{document}